\documentstyle[12pt]{article}
\title{ Dimensional Reduction and a
non-perturbative generation of a magnetic mass in non-abelian gauge theory
at Finite temperature}

\author{Satchidananda  Naik
\\
Mehta Research Institute of
 Mathematics \\
 and Mathematical Physics \\
10 Kasturba Gandhi Marg \\
(Old Kutchery Road, Katra)\\
Allahabad-211 002, INDIA\\
 e-mail: naik @ mri.ernet.in}

\begin{document}
\maketitle

\newcommand{\bee}{\begin{equation}}
\newcommand{\nn}{\nonumber}
\newcommand{\ee}{\end{equation}}
\newcommand{\ba}{\begin{array}}
\newcommand{\ea}{\end{array}}
\newcommand{\bea}{\begin{eqnarray}}
\newcommand{\eea}{\end{eqnarray}}
\newcommand{\ki}{\chi}
\newcommand{\eps}{\epsilon}
\newcommand{\pa}{\partial}
\newcommand{\lb}{\lbrack}
\newcommand{\Se}{S_{\rm eff}}
\newcommand{\rb}{\rbrack}
\newcommand{\de}{\delta}
\newcommand{\th}{\theta}
\newcommand{\rh}{\rho}
\newcommand{\ka}{\kappa}
\newcommand{\al}{\alpha}
\newcommand{\bt}{\beta}
\newcommand{\si}{\sigma}
\newcommand{\bsi}{\Sigma}
\newcommand{\vp}{\varphi}
\newcommand{\g}{\gamma}
\newcommand{\gb}{\Gamma}
\newcommand{\om}{\omega}
\newcommand{\et}{\eta}
\newcommand{\gt}{ {g^2 T }\over{4 {\pi}^2}}
\newcommand{\qab}{{{\sum}_{a\neq b}}{{q_a q_b}\over{R_{ab}}}}
\newcommand{\omb}{\Omega}
\newcommand{\pr}{\prime}
\newcommand{\ra}{\rightarrow}
\newcommand{\nb}{\nabla}
\newcommand{\MSb}{{\overline {\rm MS}}}
\newcommand{\lnh}{\ln(h^2/\Lambda^2)}
\newcommand{\df}{\delta f(h)}
\newcommand{\h}{{1\over2}}
\newcommand{\R}{m/\Lambda}
\newcommand{\abschnitt}[1]{\par \noindent {\large {\bf {#1}}} \par}
\newcommand{\subabschnitt}[1]{\par \noindent
                                          {\normalsize {\it {#1}}} \par}
\newcommand{\skipp}[1]{\mbox{\hspace{#1 ex}}}

%
%
%
%
\newcommand\dsl{\,\raise.15ex\hbox{/}\mkern-13.5mu D}
\newcommand\delsl{\raise.15ex\hbox{/}\kern-.57em\partial}
\newcommand\Ksl{\hbox{/\kern-.6000em\rm K}}
\newcommand\Asl{\hbox{/\kern-.6500em \rm A}}
\newcommand\Dsl{\hbox{/\kern-.6000em\rm D}} 
\newcommand\Qsl{\hbox{/\kern-.6000em\rm Q}}
\newcommand\gradsl{\hbox{/\kern-.6500em$\nabla$}}

\newpage
\begin{abstract} \normalsize
   A non-perturbative mass of $0.2719 g^2 T$ is generated for the
  magnetic sector of the $SU(2)$  gauge theory at high temperature
  due to the  condensation of Polyakov \cite{Pol} monopole and antimonopole
  which form a magnetic glue ball. String tension for the spatial wilson
  loop is calculated which is of the same order of magnitude and has same
  temperature dependence  as obtained from lattice simulation.

\end{abstract}

\newpage
\pagestyle{plain}
\setcounter{page}{1}
A non-perturbative generation of a mass term in the magnetic sector of the
nonabelian gauge fields at high temperature is absolutely necessary. The
lowest order perturbation (one loop) is not so infrared singular to
generate a mass term as in the case of the electric
 (time like vector potential $A_0$ ) sector  and also
 higher order perturbation
 breaks down beyond $O(g^6) $ as pointed out by Linde \cite{Linde}.
So far it is customary to cure the infrared problem
by resummation technique
through higher order perturbation which in the present case is quite subtle
due to the problem of  gauge invariance and in practice impossible beyond
$O(g^6)$  \cite{Linde}. So far it is not clear why this discrepancy exists
between the infrared behaviour of electric sector and spatial magnetic
sector of non-abelian gauge theory at finite temperature. More than a
decade ago a simple minded physical picture is given by Gross et al.
\cite{GPY} and also by Kapusta \cite{Kapst}, that the infrared
behaviour of the spatial sector is same as that of pure QCD
in three dimensions and
there one expects a mass gap which may be in the form of magnetic
 glue ball mass. So far this conjecture is
 taken as granted and to circumevent
 the infrared problem an infrared cutoff is taken which is
 same as the
 three dimensional  glue ball mass \cite{PT} obtained
  from the lattice simulation \cite{Latt}. It is also quite remarkable that the
  lattice simulation of the
  string tension of magnetic sector of the vector potential of
  the four dimensional pure QCD at finite temperature
  is quite comparable with the
  result   obtained from the three dimensional lattice   QCD at zero
  temperature \cite{Tep}.

  The generation of this mass is quite crucial to explain the
  baryon asymmetry of the universe in the frame work of standard model
  \cite{Bas}. The mass term for the gauge field weakens the discontinuous
  nature of the effective potential which eventually leads to a second
  order phase transition \cite{Dine}. In recent past there are
   several investigations in this regard and Buchm\"{u}ller et al \cite{Buch}
   put an upper bound on this mass beyond which the nature of transition
   will be second order which will wipe out the scenario of baryon asymmetry
   in the standard model.
    The main motivation of this present investigation is to show from
    the first principle how this magnetic mass arises non-perturbatively
    due to the condensation of electric strings and Polyakov monopoles
    \cite{Pol}. We present an analytic expression of this mass and
     the string tension and compare this with the lattice simulation results.

     The finite temperature partition function for the $SU(N)$ gauge theory
     is given by
     \bee
     Z(\bt)~=~\int DA_{\mu}~\exp - {{\int}_{0}^{\bt}} \int d^3x \h
     Tr F^{\mu \nu} F_{\mu \nu}.
     \ee
     The gauge fields $A_{\mu}(x, \tau)$ are periodic in the euclidian
     time $\tau$ of period $\bt={1\over T}$, where T is identified as
     temperature. Due to the periodic boundary condition
     \bee
     A_{\mu}(x,\tau)= A_{\mu}^{st}({\bf x}) +{\sum}^{ \pr \infty}_{n
     =  -\infty}A^{nst}_{\mu ,n} ({\bf x})\exp (i{\omega}_n \tau),
     \ee
     where ${\omega}_n = 2\pi nT$. The integration over the non-static
     mode $A_{\mu ,n}({\bf x})$ gives rise to an effective action
     \cite{Dired},
     \bee
    \Se = \int d^3x ~ \h  Tr F^{ij} F_{ ij} + Tr (D_{i} A^{0})^{2} +\h
     m_D^2(T) Tr A_{0}^2 +{\lambda \over{4!}} Tr A_{0}^4 ,
     \ee
     where  $m_D^2 = {{8  g^2T^2}\over 3}$ ,
      $\lambda = {{g^4T}\over {{\pi}^2}}$
     , $D_{i} A^0~=~ {\pa }_iA^0 + g {\sqrt T} [ A_i, A^0] $ and
   $A_\mu = t^a A^{a}_{\mu}$ where $t^a$ are $SU(N)$ generators.
    Here we are more interested in $SU(2)$ case because of its
    current interest in the study of the phase transitionof the early
    universe. However our theory could be extended to any gauge
    group.
   We can recognize (3)
   as Georgi-Glashow model in three dimension
   where the gauge field is
  coupled with the Higgs field $A^0$ in their adjoint representation.
  As mentioned earlier  the gauge invariant operator in
  the compact dimension (the temporal Polyakov line)
   $trP {e}^{i{\int}_{0}^{\bt} d\tau A_{0}({\bf x})}$ gets condensed in
    the vacuum. To integrate over $A^0({\bf x})$ we diagonalize
\bee
A^0({\bf x}) = \h {\big (} \ba {cc} {\vp }_3 & {\vp }_1 - i {\vp }_2 \\
{\vp }_1 + i {\vp }_2  &  -{\vp }_3 \ea {\big )} \\
 = U \phi U^{-1}
\ee
where  $ \phi = {\big (}  \ba {cc} \vp  & ~\\ ~~  & -\vp  \ea {\big )} $
and $\vp =  \mp \h \sqrt{ {\vp }^2_{1} +{\vp }_{2}^{2} +
 {\vp }^{2}_{3}}$, are
the eigenvalues of $A^0$ and $U$ is the unitary matrix which has diagonalized
it. Without any loss of generality the  gauge field is rotated
 as $A^{\pr}_{i}~=~U A_{i} U^{-1}
+i{\pa}_{i} U U^{-1} $. Thus
 \bee
   \Se = \int d^3X [({1\over{2g^2 T}}  Tr F^2_{ij}) + g^2 T
    ~ {\vp}^2 ~Tr [ A_i , {\si}^3 ]^2
    + 2 ({\pa}_i \vp )^2 + V(\vp )],
 \ee
   and the measure
   \bee
   [ {\Pi}_{X} D A_{0}({\bf x})]^{N_{\tau}} = [ {\Pi}_{X} {\vp }^2
   d\vp ]^{N_{\tau}} [dUU^{-1}]^{N_{\tau}} .
   \ee
   Here $N_{\tau}$ is the number of discrete points in the $\tau$ direction
   which gives $ N_{\tau} a = \bt = {1\over T} $ where $a$ is the lattice
   spacing. The integration over the unitary matrix gives identity
   and the rest goes to the effective action as
   \bee
   S_{measure} = -{1\over T} \int {{d^4k}\over {{(2\pi)}^4}}\int d^3X ln~
   {\vp }^2,
\ee
Here as usual  we take
$ {\sum}_{X, \tau}~ =~ {\de}^4(x=0) \int d\tau \int d^3x
 =~ \int {{d^4k}\over {{(2 \pi)}^4}}  \int d\tau \int d^3 x $. In the classical
low energy limit of the three dimensional
 effective theory $k^2 \leq m_D^2$, the Debye mass of $A_0$, taken as the
 momentum cutoff, which is the only scale parameter available. This parameter
 is not a constant which one will think it as an infrared cutoff rather this
 is an ultraviolet cutoff which depends on temperature T  and $g^2(T)$
 which is the four dimensional running coupling at scale T.  Thus
\bee
  {1\over T}   {\int}_{0}^{m_D^2} {{d^4k}\over {{(2 \pi)}^4}}
= {{ 2 g^4 T^3}\over {9 {\pi}^2}} = \al .
\ee
(This  is obtained by assuming sherical geometry of the phase space. However
this factor will not drastically change in cartesian coordinates.)
So the effective potential is
\bee
V(\vp ) =  {m_D}^2 {\vp}^2 + {{\lambda}\over 12} {\vp}^4 - \al~ln~{\vp}^2
\ee
The minimum of the potential gives
\bee
{\vp}^2_{c} = < {\vp}^2 > \approx {{g^2 T}\over {12 {\pi}^2}}
\ee
which eventually gives (c.f. eq.(5))   $m^{2}_{A^{\mp}_{i}}
= {{g^4 T^2}\over {3 {\pi}^2}}$.
This is one of our main results.

However $A^3_i$ remains massless
and by perturbation    we can never generate  a mass for this.
 We exactly follow
the path of Polyakov \cite{Pol} and show that the mass gap of the theory
is due to monopole-antimonopole condensation. In the sequel
we only present the relevant equations instead of more involving
calculations.

The non-trivial classical solutions are the monopole solutions
\bea
\vp (r) = &  {{x^3}\over r}.{\vp}_c  \nn \\
  A^3_{i} = & - {\eps}_{i3j} {{x_j}\over {r^2}}
   \eea
   for $r \geq {1\over{m_{A^{\mp}}}}$, which gives
   \bee
   \Se = {{4\pi m_{A^{\mp}}}\over {g^2 T}}
   . \ee
    For large separations of monopoles and anti-monopoles ,
    the solutions are $ \vp ={\vp}_c$ and $A^{\mp} = 0$ and
    the magnetic field due to $A^3_i$ is    that of the Dirac string
    solutions given by
    \bee
    {\bf B} = {\bf \nb \times A^3} = \h {{\bf x}\over{{\mid x \mid}^3}}-2\pi
    {\hat{e}}_{3}~ {\th}(x_3)~ \de (x_1)~ \de (x_2).
     \ee
By appropriately taking zero modes in to account for a gas of monopoles
and anti-monopoles the partition function is given by
\bee
Z = Z_{ Gaussian} Z_{monopole}
\ee
where  $ Z_{Gaussian}$ is the contribution due to
the Gaussian fluctuations around the classical  monopole solutions
      \bea
   A_i =& A_i^{cl} + {\al}_i  ,\nn \\
   \vp = &{\vp}^{cl} +\et ,  \nn\\
    and  \nn \\
   V(\vp )=& V({\vp}_{cl}) +\h V^{\pr \pr}({\vp}_{cl} ){\et}^2.
   \eea
    The classical solutions are so non-linear that the
    quadratic  fluctuations never has any effect on the low energy
    quantities namely the mass gap or the string tension of the theory,
    so we are not concerned about the fluctuations here.

\bee
Z_{monopole} = {{\sum}_{\{{q_{a}}\}}}{ {{\xi}^N}\over{N!}} \int {\prod}^{N}_{i}
  d{R}_i  \exp - ({{\pi}\over{2g^2T}} {\qab})
\ee
where  $q_a$ is the charge of the monopole sitting at $R_a$
which happened to be $\mp 1$ here,
and
\bee
\xi = {\sqrt{4\pi}}~ {{m^{7\over 2}_{A^{\mp}}}\over{g \sqrt{T}}}~
\exp (-{{4\pi m_{A^{\mp}}}\over{g^2 T}})
\ee
is the  tree level  contribution of a single monopole.
By usual Gaussian trick
\bee
Z_{monopole} = \int D\ki \exp -{\gt} \int
   d^3x  [ {(\nb \ki)}^2 - M^2 \cos {\ki}]
\ee
where
\bea
 M^2=& {8{\pi}^2 \xi }/ {g^2 T}  \nn \\
    =& e^{- {{4\over{\sqrt 3}}}}~ { 16 \over{3\pi}}~
     {1\over {{3}^{3\over 4}}}~ g^4 T^2
 \eea
 which gives $ M~=~ 0.2719~g^2 T$.
 This shows that although we start with the  massless vector field $A^3_{i}$,
 we have a massive scalar particle of mass $M$ in the vacuum. This massive
 scalar field obeying a non-linear Debye equation mediates a coloumb type of
 force between monopole and anti-monopole to form a magnetic glue ball.
 This theory has no infrared problem and the mass is also of the correct order
 of magnitude and temperature dependence which is recently taken for doing
 perturbative calculation of the effective potential to study the
 baryon asymmetry of the Universe.

    We calculate here the string tension in this formalism and compare
    our result with the  results of lattice simulation \cite{Latt}.
    \bea
    <W(C)> = & <e^{i {\oint}_C d{\xi}^i A_i({\bf x})}> \nn \\
    =& e^{- \si A}
     \eea
    where $\si $ is the string tension and A is the area of the
    Wilson loop enclosed by the contour $C$. Also
    \bee
    {\oint}_C A_i d{\xi}^i = {\int}_A {\bf B}. \hat{n} dS .
    \nn \ee
    In the semiclassical approximation \cite{Pol}, $B_i$ is expressed
    in terms of the monopole charge density as
    \bee
    B_i(x) = \h \int d^3y {{(x-y)_i}\over{\mid x-y\mid}^3}~ \rh (y)
    \ee
    where
    \bee
    \rh (y) ={\sum}_a q_a \de (y-y_a)
    \ee

    Also one can express
    \bee
    \int {\bf B}.\hat{n} dS = \int d^3x~ \et (x)~ \rh (x)
    \ee
    where
    \bee
    \et (x) = \h \int d{\bf S}_y. {{({\bf x-y})}\over{{\mid x - y \mid}^3}}
     .\ee
    Thus $ <W(C)>~ = {{Z(\et)}\over {Z(0)}}$, where
    \bee
    Z[\et ] = \int D\ki \exp -{\gt} \int d^3x [ (\nb (\ki -\et))^2 - M^2
    \cos \ki ].
    \ee
    To   conveniently  evaluate the string tension $\si$ ,
     without any loss of generality
     we chose the contour $C$ of the Wilson Loop as
      a circle of radius ${1\over M}$ lying in the X-Y plane.
      The  ansatz for the classical solution of
      $\ki $ field is chosen as one dimensional Sine-Gordon
      kink in the transverse
      z   direction which is given as
     \bea
     {\ki }_{cl}(z) =& 4 \arctan (e^{-mz})~~ z >0  \nn \\
     {\ki }_{cl} (-z) =& - {\ki }_{cl}(z).
      \eea

     Substituting this  we get $ \si = M g^2T/4$ which gives
     $\sqrt{\si} = 0.266~ g^2 T$. The string tension of spatial Wilson
     loop of 4D SU(2) lattice gauge theory at finite temperature is
      $\sqrt{\si} =  0.369 g^2 T$    \cite{Latt}.

     The main observation of this analysis is that   {\it there is
    no infrared problem for gauge theory which looks completely
    maligned with this to start with.} The one-loop perturbative
    contribution of the non-static modes, which gives Debye mass
     to $A^0$  explicitly,
    however non-perturbatively   generate massive excitation in
    the three dimensional vacuum. The most striking feature of
    our endeavour is the agreement of our string tension with
    correct order of magnitude and the temperature dependence
    which is obtained by lattice simulation.

  {\it Acknowledgements}

     It is a pleasure to thank Professor H.S. Mani, S. Kalyan Rama
     and all the members of HEP group  for
      helpful discussion and encouragement. Thanks are also
      due to Professors Ashok Das,
      Ashoke Sen and Jan Ambjorn for their  comments and suggestions and
      all the   members of Niels Bohr Institute for their
       hospitality where a part of this work was done.


\newpage
   \begin{thebibliography}{99}

   \bibitem{Pol}
   A.M. Polyakov, Nucl Phys B 120  429 (1977),
   \\ Phys. Lett. B 72 477 (1977).
   %
    \bibitem{Linde}
   %
   A  .D. Linde, Phys. Lett. B 96 289 (1980).
  %
   \bibitem{GPY}
   %
   D.Gross, R.D. Pisarski and L.G. Yaffe,
   \\ Rev. Mod.Phys. 53 43 (1984).
   %
   \bibitem{Kapst}

    J.Kapusta, Finite temperature Field Theory,
    \\  Cambridge university Press,
         New York (1990).
   %
  \bibitem{PT}
  K. Farakos et al. Nucl. Phys. B 245 67 (1994),\\
  M. Quiros, J.R. Espinosa and F. Jwirner,
  \\ Phys Lett B 314 206 (1993),\\
   W. Buchm\"{u}ller, T. Heilbig and D. Walliser,
    \\ Nucl.Phys.B 407 387 (1993), \\
   M. Reueter and C. Wetterich, Nucl. Phys. B 408  91 (1993).
   %
   \bibitem{Latt}
   G.Bali et al. Phys.Rev. Lett 71 3059 (1993), \\
   J.Fingberg et al. Nucl. Phys B (proc. Supp)34 295 (1994), \\
   Frithjof Karsch, Lattice 93
   \\ Nucl. Phys. B (proc. Supp) 34 63 (1994).
   \bibitem{Tep}
   M. Teper, Phys.Lett B 289  115 (1992).
    %
   \bibitem{Bas}
    %
   V.A. Kuzmin, V.A. Rubakov and M.E. Shaposhnikov,\\
   Phys.Lett B 155  36 (1995),\\
    M.E. Shaposhnikov, Nucl.Phys.B 287 757 (1987), \\
   M. Dine et al. Phys. Lett B 283 219  (1993), \\
   J. Ambjorn et al. phys.Lett. B 244 477  (1990),\\
   J. Ambjorn and K. Farakos,  Phys. Lett. B 294 248 (1992),\\
   L. Mclerran et al. Phys. Lett B 256 451 (1990).
    \bibitem{Dine}
    %
    M. Dine, et al. Phys. Rev. D 46 550  (1992),\\
    F. Karsch,T. Neuhaus and A. Patkos,
    \\ University Of Bielefeld Report-94-27.
      \bibitem{Buch}
      W. Buchm\"{u}ller et al. , Annals of Phys. 234 260 (1994).
   \bibitem{Dired}
   S. Nadkarni, phys. Rev D 27 917(1983),
   \\  Phys. Rev. Lett. 61 396 (1988) ,\\
    T. Reisz, Z  Phys. C 53 169 (1992),  \\
   N.P. Landsman, Nucl. Phys. B 322  498 (1989),\\
   L. K\"{a} rK\" {a}inen et al. Nucl.Phys. B 395 733(1993).

\end{thebibliography}
\end{document}